\documentclass{iaus}
\usepackage{graphicx,natbib}
\newcommand{\be}[1]{\begin{equation}\label{#1}}
\newcommand{\ee}{\end{equation}}
\newcommand{\bea}[1]{\begin{eqnarray}\label{#1}}
\newcommand{\eea}{\end{eqnarray}}  

\newcommand{\dxdy}[2]{\frac{ \partial #1 }{ \partial #2 }}

\renewcommand{\eg}  {\it e.g.}        % exempli gratia {f.eks.}
\newcommand{\cf}  {\it cf.}         % confer {jvnf.}
\newcommand{\ie}  {\it i.e.}        % id est {d.v.s.}

\font\mib=cmmib10 scaled \magstep0

\title[3D-simulation of A-star convection] %% give here short title %%
{3D-simulation of the Outer Convection-zone of an A-star}

\author[R. Trampedach]   %% give here short author list %%
{Regner Trampedach$^1$}

\affiliation{$^1$Research School of Astronomy and Astrophysics,
			 Mt.\ Stromlo Observatory,Cotter Road, Weston ACT 2611, Australia
			 \break email: art@mso.anu.edu.au}

\pubyear{2004}
\volume{224}  %% insert here IAU Symposium No.
\pagerange{119--126}
\date{?? and in revised form ??}
\jname{The A-Star Puzzle}
\editors{J.\,Zverko, W.W.\,Weiss, J.\,\v{Z}i\v{z}\v{n}ovsk\'{y}, \& S.J.\,Adelman, eds.}
\begin{document}

\maketitle

%%%%%%%%%%%%%%%%%%%%%%%%%%%%%%%%%%%%%%%%%%%%%%%%%%%%%%%%%%%%%%%%%%%%%%%%%%%
\begin{abstract}
The convection code of Nordlund \& Stein has been used to evaluate the
3D, radiation-coupled convection in a stellar atmosphere with
$T_{\rm eff} = 7300$\,K, log$g = 4.3$ and [Fe/H]$ = 0.0$, corresponding to
a main-sequence A9-star.
I will present preliminary comparisons between the 3D-simulation
and a conventional 1D stellar structure calculation, and elaborate on the
consequences of the differences.
\end{abstract}
%%%%%%%%%%%%%%%%%%%%%%%%%%%%%%%%%%%%%%%%%%%%%%%%%%%%%%%%%%%%%%%%%%%%%%%%%%%

\firstsection % if your document starts with a section,

%%%%%%%%%%%%%%%%%%%%%%%%%%%%%%%%%%%%%%%%%%%%%%%%%%%%%%%%%%%%%%%%%%%%%%%%%%%
\section{Introduction}
%%%%%%%%%%%%%%%%%%%%%%%%%%%%%%%%%%%%%%%%%%%%%%%%%%%%%%%%%%%%%%%%%%%%%%%%%%%

From 3 dimensional simulations of convection, it has been known for the
last two decades that convection grows stronger with increasing effective
temperature and decreasing gravity. By stronger, is here meant larger
Mach-numbers, larger turbulent- to total-pressure ratios and larger convective
fluctuations in temperature and density. The stronger convection has also been
accompanied by increasing departures from 1D stellar models that fail to predict
the extensive overshoot into the high atmosphere, the turbulent pressure and
its effect on the hydrostatic equilibrium, the temperature fluctuations and
the coupling with the highly non-linear opacity. The latter has the effect
of heating the layers below the photosphere, thereby expanding the atmosphere,
as also done by the turbulent pressure. The various 1D convection
theories/formulations, {\eg}, classical mixing-length \citep{boehm:mlt},
non-local extensions to it \citep{gough:MLT-PulsStars} or an independent
formulation based on turbulence \citep{canuto-mazzitelli:conv-improv}, all have
similar shortcomings with respect to the simulations. Their predictive power
is further limited by the free parameters involved.

Going towards earlier type stars, apart from stronger convection, also means
a more shallow outer convection zone. This combination is rather unpredictable
and is the main motivation for the work presented here. Classical predictions
call for the outer convection zone to disappear close to the transition
between A and F stars, but details about where and how this transition
occurs can only be gained from realistic, 3D simulations, as outlined below.

%%%%%%%%%%%%%%%%%%%%%%%%%%%%%%%%%%%%%%%%%%%%%%%%%%%%%%%%%%%%%%%%%%%%%%%%%%%
\section{The simulations}
%%%%%%%%%%%%%%%%%%%%%%%%%%%%%%%%%%%%%%%%%%%%%%%%%%%%%%%%%%%%%%%%%%%%%%%%%%%
The simulation presented here was carried out using the code of
\citet{aake:comp-phys} and is further described in
\citet{aake:numsim1}, \citet{bob:conv-waves} and \citet{bob:Tuebingen}.

The basis of hydrodynamics is the Navier-Stokes equations, and the code 
employs the {\em conservative} or {\em divergence} form
\bea{NavStok1}
    \dxdy{\varrho}{t} &=& -\nabla\cdot(\varrho\hbox{\mib u}) \\
    \label{NavStok2}
	{\partial\varrho\hbox{\mib u}\over \partial t} &=&
		- \nabla\cdot(\varrho\hbox{\mib u}\hbox{\mib u}) - \nabla P_{\rm g}
		+ \varrho\hbox{\mib g}\\
    \label{NavStok3}
    \dxdy{\varrho\varepsilon}{t} &=&
       -\nabla\cdot(\varrho\varepsilon\hbox{\mib u})
	   -P_{\rm g}\nabla\cdot\hbox{\mib u} + \varrho(Q_{\rm rad}+Q_{\rm visc})\ ,
\eea
where $\varrho$ is the density, $P_{\rm g}$ is the gas pressure, $\varepsilon$
is the specific internal energy, {\mib u} is the velocity field, {\mib g} is
the gravitational acceleration and $Q_{\rm rad}$ and $Q_{\rm visc}$ are the
radiative and viscous heating, respectively, the latter arising from the
numerical diffusion applied.

Equations (\ref{NavStok1})--(\ref{NavStok3}) describe the conservation of
mass, momentum and energy, respectively, with sources and sinks on the
right-hand-side. For the convection code the equations are preconditioned, by
dividing by $\varrho$, to improve handling of the large density contrast
between the top and bottom of the simulations.

The vertical component of the momentum equation, Eq.\ (\ref{NavStok2}), can be
written
\be{Fz}
	F_z = - \dxdy{(\varrho u_z^2 + P_{\rm g})}{z} + \varrho g\ ,
\ee
as we have chosen {\mib g} to be in the $z$-direction. With $F_z=0$ this
equation describes hydrostatic equilibrium, where the gas pressure and the
turbulent pressure, $P_{\rm turb}=\varrho u_z^2$, provide support against
gravity.

The gas pressure, $P_{\rm g}(\varrho,\varepsilon)$, and the opacities going
into the computation of the radiative heating, $Q_{\rm rad}$, form the
atomic physics basis for the simulations. The continuous opacities are
calculated from the {\sc MARCS}-package \citep{b.gus} and subsequent updates
as detailed in \citet{trampedach:thesis}, the line opacity is in the form
of opacity distribution functions (ODFs) \citep{kur:line-data}, and the
equation of state accounts explicitly for all ionization stages of the 15 most
abundant elements \citep{mhd1,mhd3}.

The present simulation is performed on a $100\times 100\times 82$-point grid,
has $T_{\rm eff} = 7300$\,K, log$g = 4.3$ and [Fe/H]$ = 0.0$, and therefore
corresponds to a A9 dwarf on the main-sequence. The computational domain is
11.5\,Mm on each side, and 13.1\,Mm deep, of which 1.5\,Mm is above the
photosphere.

So far the convection code has only been used for stars that were convective
at the bottom boundary, so in order to accommodate this simulation, the
boundary condition was changed and evaluation of radiative heating in
optically thick layers was included.

The simulation was carried out in the plane-parallel approximation and
includes no rotation or magnetic fields, and has a simple Solar abundance
\citep{AG89}.
Both thermodynamics and radiative transfer is performed in strict LTE.
The velocity-field is too large, through-out the simulation domain, to
support segregation of elements, rendering diffusion and radiative levitation
of individual species, comfortably irrelevant.

%%%%%%%%%%%%%%%%%%%%%%%%%%%%%%%%%%%%%%%%%%%%%%%%%%%%%%%%%%%%%%%%%%%%%%%%%%%
\section{Comparison with 1D models} \label{sect:cmp1d}
%%%%%%%%%%%%%%%%%%%%%%%%%%%%%%%%%%%%%%%%%%%%%%%%%%%%%%%%%%%%%%%%%%%%%%%%%%%

\begin{figure}
	\includegraphics[width=14cm]{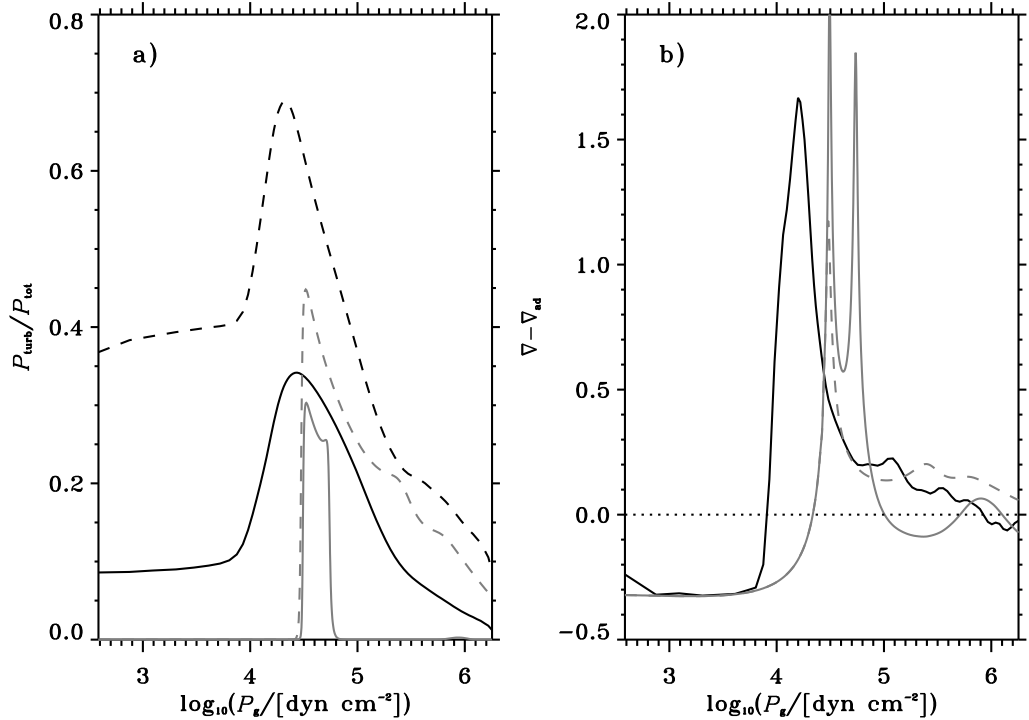}
	\caption{\label{cmp1d}Comparisons between the simulation and two
			 1D, mixing-length models. In both panels the black solid line
			 shows the temporal- and horizontally averaged simulation, and the
			 gray solid line shows the same quantity for a 1D model with
			 $\alpha=1.0$ and the dashed gray curve is for $\alpha=2.0$.
			 Panel {\bf a)} shows the turbulent- to total-pressure ratios,
			 and the dashed black line shows the RMS Mach-numbers.
			 Panel {\bf b)} shows the super-adiabatic gradient, with the
			 dotted zero-line aiding the location of the borders of the
			 convection zones.}
\end{figure}

In Fig.\ \ref{cmp1d} the temporal and horizontal average of the simulation is
compared with a corresponding 1D stellar model. From the solid black curve in
panel {\bf a)} we see that this simulation has an impressive turbulent pressure,
making up almost 35\% of the total pressure about 700\,km below the
photosphere (it is about 13\% in Solar simulations). The mach-numbers
reaching Mach 0.7 as shown with the dashed black line, is equally impressive.
The convection is much less efficient compared to the Solar case, and invokes
both higher velocities, as seen in panel a), and higher super-adiabatic
gradient, as seen from the black curve in Panel b) of Fig.\ \ref{cmp1d}.
The two gray curves show the $P_{\rm turb}/P_{\rm tot}$-ratio and
$\nabla-\nabla_{\rm ad}$ for two 1D envelope models using the standard
mixing-length formulation of convection \citep{boehm:mlt}. The solid line is
for $\alpha=1.0$ and the dashed line for $\alpha=2.0$, in an attempt to
bracket the behavior of the simulation. One glance at Fig.\ \ref{cmp1d}
makes it clear that no mixing-length model can reproduce the outer few percent
of an A9 dwarf. The convection zone extends further up into the atmosphere and
has much broader features than possible with mixing-length models. We also
see that overshoot from the convection zone supports an appreciable
velocity-field, with velocities of 2--3\,km\,s$^{-1}$, and obvious consequences
for spectral line shapes.

The large $P_{\rm turb}$ results in smaller temperatures at the same
hydrostatic pressure, when compared to a 1D model. This can be translated into
higher pressures and densities on an optical depth scale. Consequently
the derived population of line-producing states in atoms, ions
and molecules will be misleading and result in erroneous abundance analysis.

The mismatch between the simulation and the 1D models in the atmosphere is
actually so profound that no mixing-length model with $T_{\rm eff}$ and
$g_{\rm surf}$, consistent with the simulation, can be found to match the
simulation ({\ie}, $P$, $\varrho$ {\em and} $T$) at the bottom. This means
that stellar structure and evolution calculations are misplaced in the
HR-diagram, with implications for age-determinations of globular clusters
and our general knowledge of the interior and evolution of A-stars.

%%%%%%%%%%%%%%%%%%%%%%%%%%%%%%%%%%%%%%%%%%%%%%%%%%%%%%%%%%%%%%%%%%%%%%%%%%%
\section{Temperature inversion} \label{sect:Tinv}
%%%%%%%%%%%%%%%%%%%%%%%%%%%%%%%%%%%%%%%%%%%%%%%%%%%%%%%%%%%%%%%%%%%%%%%%%%%

This simulation, which is the hottest performed with this code, displays
some large and persistent temperature inversions, 1.5-2.5\,Mm below the
photosphere. They have an amplitude of up to 8\,000\,K and are about
0.5\,Mm deep. During the meeting, Noels suggested that there might be a
correlation between the turbulent pressure and the low temperatures; keeping
$P_{\rm tot}$ and $\varrho$ constant to ensure hydrostatic equilibrium, a
large $P_{\rm turb}$ would force a low temperature. This is, however not
observed in the simulations. There is no correlation between the temperature
inversion and $P_{\rm turb}$ -- nor are there correlations with vertical or
horizontal velocities,
or the density contrast. There is, however, a strong correlation
with the vertical force (see Eq.\ \ref{Fz}), as the force on the plasma in
the temperature inversion is always outwards (the $-z$-direction). Furthermore,
the vorticity is clearly skewed towards larger values in the inversion layer,
especially when compared to the upflow, but also with respect to the
down-drafts.
\begin{figure}\
	\includegraphics[width=10cm]{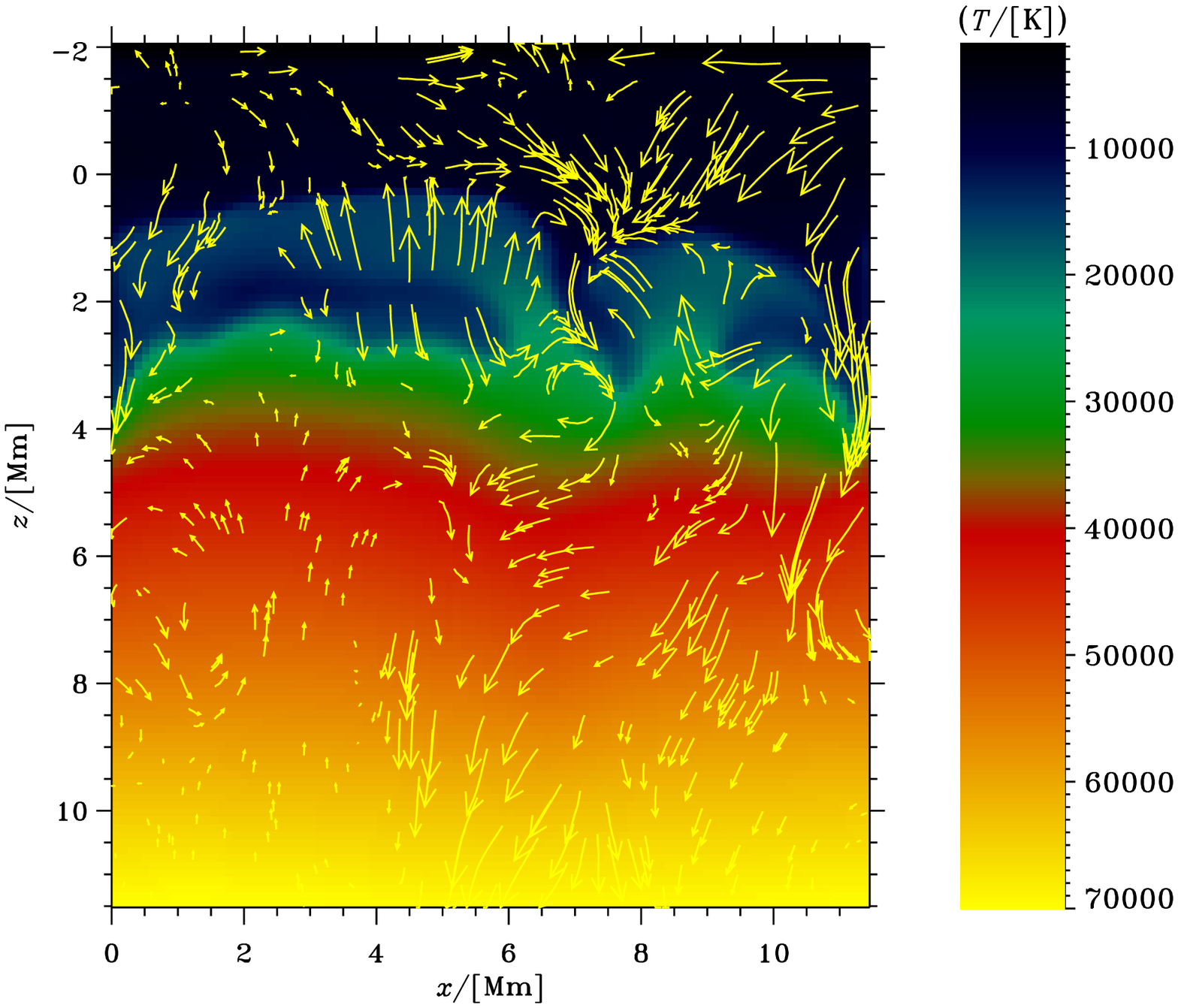}
	\caption{\label{Tinv}A vertical slice of a snapshot showing a temperature
			 inversion. The color-scale shows temperature, and the arrows
			 indicate the velocity-field.}
\end{figure}

Fig.\ \ref{Tinv} shows a vertical snapshot in temperature, displaying a 
typical temperature inversion. The arrows indicate the velocity field, with the
maximum length corresponding to 2.9\,km\,s$^{-1}$. The temperature inversion
extends some 6\,Mm across the left side of the plot, at a height of 2\,Mm.
It is connected
to the downdraft at the edge, but it never connects with the downdraft to
the right of the center. The velocity field there is connected, though, but the
colder plasma from the inversion is compressed on the way to the downdraft,
and therefore heated. The velocity field is diverging in the $z$-direction,
neutral in the $x$-direction and convergent in the $y$-direction, perpendicular
to the plane of the figure -- the net-flow is convergent.
The features observed in this  simulation are cylindrical in
the horizontal direction, with width and depth being similar and the length
being 5--10 times larger.

The temperature inversions seem to develop when the local photosphere has
subsided to about 2\,Mm below the average photosphere, at the edge of a granule.
The layers above, closer to the height of the average photosphere, then heats
up, leaving a cooler area in between -- the temperature inversion. In
white-light surface intensity, this now looks like the edge of a normal granule.
The inversion immediately starts eroding from the newly heated region on top
moving down, and presumably from heating by the surroundings. The temperature
profile after this sequence, looks like that of a downdraft, and from the
surface, what looked like the edge of a normal granule, collapse as the
temperature inversion below disappear. The whole cycle takes about 5\,min.

The reason for this behavior is still under investigation, but it might be
connected to the local minimum in $\nabla_s$, as seen in Fig.\ \ref{cmp1d}
around $\log_{10}P_{\rm g}\simeq 4.9$ corresponding to $z\simeq 1.9$\,Mm.

%%%%%%%%%%%%%%%%%%%%%%%%%%%%%%%%%%%%%%%%%%%%%%%%%%%%%%%%%%%%%%%%%%%%%%%%%%%
\section{Summary}
%%%%%%%%%%%%%%%%%%%%%%%%%%%%%%%%%%%%%%%%%%%%%%%%%%%%%%%%%%%%%%%%%%%%%%%%%%%

A realistic 3D simulation of convection in the surface layers of a A9 dwarf
reveals profound differences between that and a conventional 1D stellar
structure model, as indicated in Sect.\ \ref{sect:cmp1d}.
In general, the simulations have much smoother and broader convection-zone
features, compared to mixing-length models.

We are also treated to a new phenomena, as the simulations display repeated
local temperature inversions just below the photosphere
({\cf}, Sect.\ \ref{sect:Tinv}). The mechanism
responsible for these large inversions, have not been uncovered yet, but
it being an effect of large fluctuations in the turbulent pressure, has been
ruled out.

This is still work in progress, and in the future, this and other simulations
will be used for evaluation of spectral lines, limb-darkening, broad-band
colors, granulation-spectra and p-mode spectra.
 
%%%%%%%%%%%%%%%%%%%%%%%%%%%%%%%%%%%%%%%%%%%%%%%%%%%%%%%%%%%%%%%%%%%%%%%%%%%
%\bibliography{bibs/eos,bibs/opac,bibs/seism,bibs/conv,bibs/staratm,bibs/starmod}

%%%%%%%%%%%%%%%%%%%%%%%%%%%%%%%%%%%%%%%%%%%%%%%%%%%%%%%%%%%%%%%%%%%%%%%%%%%

\begin{discussion}

\discuss{Cowley}{Did you get a higher contrast between hot and cold regions
than in simulations of cooler stars? At one time I though the ionization
imbalance that we observe in some cool Ap stars might be explained by such
temperature inhomogeneities. But I did some test calculations and found this
could not explain the ionization -- at least not with the numerical models
I had.}
\discuss{Trampedach}{The RMS temperature fluctuations at $\tau=1$ on a
Rosseland optical depth scale, are about twice as large for the A9 simulation
($\sim 610$\,K) as for the Solar simulation ($\sim 300$\,K). For the latter,
the RMS fluctuations stay below 750\,K, peaking at $\tau=10$, whereas in the
A9 simulation they grow almost exponentially to reach 5000\,K at $\tau=100$.

I don't know what kind of
numerical models you have used, but a decisive factor in the emergent spectra
from the convection simulations is the asymmetry of the distributions of
temperatures, densities and velocities, as well as their correlations.
These characteristics are hard to predict without realistic simulations.}

\discuss{Piskunov}{What is the height of the temperature inversion and the
filling factor of the effect? Do you see significant changes of the filling
factor?}
\discuss{Trampedach}{The temperature inversions are about 0.5--1.0\,Mm deep and
peak at a depths of 1.5--2.5\,Mm. They involves from 2\% to 8\% of the
horizontal area, with an average of about 5\%.}

\discuss{Noels}{Do you find a high turbulent pressure in the region of
temperature inversion? If so, I would suggest to investigate this point
in relation to the origin of the temperature inversion.}
\discuss{Trampedach}{There is no obvious correlation between the temperature
inversion and the local turbulent pressure. I would refer to
Sect.\ \ref{sect:Tinv} for the present extent of my analysis.}

\discuss{Grevesse}{Micro-turbulence and macro-turbulence needed with 1D
stellar photosphere models are not needed anymore with 3D models. Why is it
not the same when 3D models are used, instead of 1D models, for A stars?}
\discuss{Trampedach}{I don't think that has changed, going from solar-like
stars, to A stars. I think what you are alluding to, is the lack of reversal of
bisector shape in the A star simulations presented by Freytag (elsewhere in
these proceedings). I don't know the reason for this and the only major thing
missing from those simulations is line-blanketing. Whether that will make
the difference is unclear and will have to await further simulations.
It is important to understand that theoretical bisectors from 1D models are
straight, no matter how much micro- or macro-turbulence is applied. The
shape of bisectors is a higher order problem, that cannot be addressed with
1D models.
}

\end{discussion}
%%%%%%%%%%%%%%%%%%%%%%%%%%%%%%%%%%%%%%%%%%%%%%%%%%%%%%%%%%%%%%%%%%%%%%%%%%%

\end{document}